\documentclass[pra,final,twocolumn,showpacs,superscriptaddress,aps]{revtex4-1}
\usepackage{graphicx}
\usepackage{lipsum}
\usepackage{amssymb}
\usepackage{amsmath}
\usepackage{color}
\usepackage{multirow}
\usepackage{afterpage}
\usepackage{dsfont}

\begin{document}
\title{Fidelity Based Measurement Induced Nonlocality}

\author{R. Muthuganesan and R. Sankaranarayanan}
\affiliation{Department of Physics, National Institute of Technology, Tiruchirappalli - 620015, TamilNadu, India.}

\begin{abstract}
In this paper we propose measurement induced nonlocality (MIN) using a metric based on fidelity to capture global nonlocal effect of a quantum state due to locally invariant projective measurements. This quantity is a remedy for local ancilla problem in the original definition of MIN. We present an analytical expression of the proposed version of MIN for pure bipartite states and $2\times n$ dimensional mixed states. We also provide an upper bound of the MIN for general mixed state. Finally, we compare this quantity with MINs based on Hilbert-Schmidt norm and skew information for higher dimensional Werner and isotropic states.
\end{abstract}

\maketitle

\section{Introduction}
One of the most counter-intuitive features of quantum mechanics is its nonlocal nature, which leads to fundamental departure from classical physics. Over the years, investigation on nonlocality of a quantum system has been centred around entanglement, non-classical correlation between different parts of a composite system -a valuable resource for various information processing task \cite{Bennett1993, Bouwmeester1997, Bennett1992, Ekert1991, Shor1995}. Since the pioneering work of Bell \cite{Bell1964}, entanglement is believed to be the only manifestation of quantum nonlocality. In other words, entangled states are beyond the purview of local hidden variable model and hence violate Bell's inequality. Further exploration of composite system revealed the intriguing complexity in quantum states. In particular, Werner showed that while all pure entangled states violate Bell's inequality, \textit{all} mixed entangled states do \textit{not} violate the inequality \cite{Werner1995}. This is attributed to the presence of noise or mixedness, which are responsible in destroying nonlocal correlation between different parts of the composite system, and hence some of the mixed entangled states behave locally \cite{Almeida2007}. It is now broadly accepted that entanglement is \textit{not} the complete manifestation of nonlocality.

In light of this, better quantification of quantum nonlocality is instructive to reveal the complexity of composite states. Recently, Luo and Fu presented a new measure of nonlocality
for bipartite system in the perspective of measurements, termed as measurement induced nonlocality (MIN) \cite{Luo2011}. This quantity is in a sense complementary to the geometric measure of quantum discord \cite{Dakic2010}. In other words, MIN can quantify the nonlocal resource in quantum communication protocols involving local measurement. It is worth noting that pre- and post-measurement states and found to be useful in some of the information processing like quantum dense coding, remote state control and quantum state steering \cite{Schrodinger1936,Wiseman2007, Peters2005, Mattle1996, Li2002, Xiang2005}.

MIN characterizes nonlocality of a quantum state in the perspective of locally invariant projective measurements, and hence more general than the Bell nonlocality. One important merit of this quantity is that it can be evaluated analytically for any $2\times n$-dimensional state. However, there is a problem with this geometric (Hilbert-Schmidt norm) MIN that it may change rather arbitrarily through some trivial and uncorrelated action of the unmeasured party - local ancilla problem. As shown elsewhere \cite{Chang2013}, this issue can be resolved by replacing density matrix with its square root. MIN has also been investigated based on relative entropy \cite{Xi2012}, von Neumann entropy \cite{Hu2012}, skew information \cite{Li2016} and trace distance \cite{Hu2015}. Further, MIN has been investigated for bound entangled states \cite{Rana2013}, general bipartite system \cite{Mirafzali2011} and Heisenberg spin chains \cite{Chen2015, Muthuganesan2016}. The dynamics and monogamy of measurement induced nonlocality also has been studied \cite{Sen2012, Sen2013}.

In this article, we propose the MIN based on fidelity induced metric. It is shown that this quantity is remedying the local ancilla problem associated with the geometric MIN and also easy to measure. We derive an analytical expression of fidelity based MIN for pure state, which coincides with the geometric MIN (discord). Further, we provide an upper bound for arbitrary $m\times n$ dimensional mixed state and a closed formula for $2\times n$ dimensional mixed state. The new version of MIN is also shown to be consistent with other forms of MIN for two well-known families of states, namely Werner and isotropic states.

\section{MIN based on Fidelity}
Fidelity is a measure of closeness between two arbitrary states
$\rho $ and $\sigma $, defined as $F(\rho,\sigma )= \left(tr\sqrt{\sqrt{\rho }\sigma \sqrt{\rho }} \right)^{2}$ \cite{Jozsa1994}. This measure has been explored in various context of quantum information processing such as cloning \cite{Gisin1997}, teleportation \cite{Zhang2007}, quantum state tomography \cite{Bogdanov2010}, quantum chaos \cite{Gorin2006} and spotlighting phase transition in physical systems \cite{Gu2010}. Though fidelity itself is not a metric, one can define a metric $D(\rho,\sigma )=\Phi (F(\rho,\sigma))$, where $\Phi$ is a monotonically decreasing function of $F$, and is required to satisfy all the axioms of distance measure. Few such fidelity induced metrics are Bures angle $A(\rho,\sigma )=arccos\sqrt{F(\rho,\sigma )}$, Bures metric $B(\rho,\sigma )=\left(2-2 \sqrt{F(\rho,\sigma )}\right)^{1/2}$ and sine metric $C(\rho,\sigma )=\sqrt{1-F(\rho,\sigma )}$\cite{Gilchrist2005}.

Since the computation of fidelity involves square root of density matrix, various forms of fidelity have been proposed to ease the computation. Here we follow one such form \cite{Wang2008}.
\begin{eqnarray}
\mathcal {F}(\rho ,\sigma )=\frac{(tr(\rho \sigma ))^{2}}{tr(\rho)^{2}tr(\sigma )^{2}}
\end{eqnarray}
to define a metric as $\mathcal {C}(\rho ,\sigma )=\sqrt{1-\mathcal {F}(\rho ,\sigma )}$.

Let us consider a bipartite quantum state $\rho $ shared by the parties $a$ and $b$ with respective system state spaces $\mathcal{H}^{a}$ and $\mathcal{H}^{b}$. Defining MIN in terms of fidelity induced metric as
\begin{align}
 N_{\mathcal {F}}(\rho ) =~^{max}_{\Pi ^{a}}\mathcal {C}^{2}(\rho, \Pi ^{a}(\rho )) \label{min}  
\end{align}
where the maximum is taken over the von Neumann projective measurement on subsystem $a$.
Here $\Pi^{a}(\rho) = \sum _{k} (\Pi ^{a}_{k} \otimes   \mathds{1} ^{b}) \rho (\Pi ^{a}_{k} \otimes    \mathds{1}^{b} )$, with $\Pi ^{a}= \{\Pi ^{a}_{k}\}= \{|k\rangle \langle k|\}$ being the projective measurements on the subsystem $a$, which do not change the marginal state $\rho^{a}$ locally i.e., $\Pi ^{a}(\rho^{a})=\rho ^{a}$. In other words, MIN is defined in terms of the fidelity between pre- and post-measurement state. Here we list out some interesting properties of the MIN as defined above.
\begin{enumerate}
\item[(i)]  $N_{\mathcal {F}}(\rho )$ is a positive quantity i.e., $N_{\mathcal {F}}(\rho )\geq 0.$
\item[(ii)] $N_{\mathcal {F}}(\rho )=0$ for any product state $\rho=\rho_{a}\otimes  \rho _{b}$ and the classical state in the form $\rho =\sum _{i}p_{i}|i\rangle \langle i| \otimes   \rho_{i}  $ with nondegenerate marginal state $\rho^{a}=\sum_{i}p_{i}|i\rangle \langle i|   $.
\item[(iii)] $N_{\mathcal {F}}(\rho )$ is locally unitary  invariant in the sense that $N_{\mathcal {F}}\left((U\otimes   V)\rho  (U\otimes   V)^\dagger\right)=N_{\mathcal {F}} (\rho) $ for any unitary operators $U$ and $V$.
\item[(iv)] For any pure maximally entangled state $N_{\mathcal {F}}(\rho) $ has the maximal value of $0.5$. 
\item[(v)] $N_{\mathcal {F}}(\rho )$ is invariant under the addition of any local ancilla to the unmeasured party (proof
is given below). 
\end{enumerate}

Originally MIN is defined as the square of Hilbert-Schmidt norm of difference of pre- and post-measurement state i.e., \cite{Luo2011}
\begin{align}
 N(\rho ) =~^{max}_{\Pi ^{a}}\| \rho - \Pi ^{a}(\rho )\| ^{2} 
\end{align}
where the maximum is taken over all local projective measurements. One problem of this geometric MIN is that it may change rather arbitrarily through some trivial and uncorrelated action on the unmeasured party $b$. This arises from appending an uncorrelated ancilla $c$ and regarding the state $\rho ^{a:bc}=\rho^{ab} \otimes  \rho ^{c}$ as a bipartite state with the partition $a:bc$; then
\begin{align}
 N(\rho^{a:bc} ) = N(\rho^{ab}) tr (\rho ^{c})^2  \nonumber
\end{align}
implying that MIN differs arbitrarily due to local ancilla $c$  as long as $\rho^{c}$ is mixed. This problem of MIN can be circumvented with the fidelity based MIN as defined in eq. (\ref{min}). After the addition of local ancilla the fidelity between the pre-and post-measurement state is
\begin{align}
 \mathcal {F}\left(\rho^{a:bc},\Pi ^{a}(\rho^{a:bc})\right) = \mathcal {F}\left(\rho^{ab}\otimes  \rho ^{c},\Pi ^{a}(\rho^{ab})\otimes  \rho ^{c}\right). \nonumber
\end{align}
Using multiplicativity property of fidelity \cite{Jozsa1994},
\begin{align}
 \mathcal {F}\left(\rho^{a:bc},\Pi ^{a}(\rho^{a:bc})\right) =& \mathcal {F}\left(\rho^{ab}, \Pi ^{a} (\rho ^{ab})\right).\mathcal {F}(\rho^{c},\rho^{c})\nonumber \\
 =&\mathcal {F}\left(\rho^{ab},\Pi ^{a}(\rho^{ab})\right)  \nonumber
\end{align}
resulting the property (v) of the fidelity based MIN. Hence $N_{\mathcal {F}}(\rho)$ is a good measure of nonlocality or quantumness in a given system. 
\section{MIN for Pure state}
{\bf Theorem 1:} For any pure bipartite state with Schmidt decomposition $| \Psi \rangle =\sum_{i}\sqrt{\lambda _{i}}| \alpha _{i} \rangle \otimes | \beta _{i}\rangle $,
\begin{align}
 N_{\mathcal {F}}(| \Psi \rangle\langle \Psi| )=1- \sum_{i}\lambda_{i}^{2}.
\end{align}
The proof is as follows. The von Neumann projective measurement on party $a$ is expressed as $\Pi ^{a}=\{\Pi ^{a}_{k}\} = \{U| \alpha_{k}\rangle \langle \alpha_{k}| U^{\dagger}\} $ for any unitary operator $U$. The projective measurements do not alter the marginal states i.e., $\left(\Pi ^{a}(\rho^{a})=\sum_{k}\Pi ^{a}_{k}\rho^{a}\Pi ^{a}_{k}=\rho^{a}\right)$. In general,
\begin{align}
 \rho^{a}= \sum_{k} U| \alpha_{k}\rangle \langle \alpha_{k}| U^{\dagger}\rho^{a} U| \alpha_{k}\rangle \langle \alpha_{k}| U^{\dagger}.  \nonumber
\end{align}
This marginal state $\rho^{a}$ can be written as spectral decomposition in the orthonormal bases $\{U| \alpha_{k}\rangle\} $ as
 \begin{align}
 \rho^{a}= \sum_{k} \langle \alpha_{k}| U^{\dagger}\rho^{a} U| \alpha_{k}\rangle U| \alpha_{k}\rangle \langle \alpha_{k}| U^{\dagger} 
\end{align}
 where $\langle \alpha_{k}| U^{\dagger}\rho^{a} U| \alpha_{k}\rangle =\lambda_{k}$, the eigenvalues of $\rho^{a}$. Since $\rho=| \Psi \rangle \langle \Psi| = \sum_{ij}\sqrt{\lambda_{i}\lambda_{j}}| \alpha_{i} \rangle \langle \alpha_{j}| \otimes  | \beta_{i} \rangle \langle \beta_{j}|$, $\Pi ^{a}(\rho)=\sum_{k}(\Pi ^{a}_{k}\otimes  \mathds{1} )\rho(\Pi ^{a}_{k}\otimes  \mathds{1} )$ becomes
 \begin{align}
 \Pi ^{a}(\rho) &=\sum_{k}\sum_{ij}\sqrt{\lambda_{i}\lambda_{j}}\langle \alpha_{k}| U^{\dagger}|\alpha_{i}\rangle \langle \alpha_{j}| U|\alpha_{k}\rangle \nonumber \\
& U| \alpha_{k}\rangle \langle  \alpha_{k}| U^{\dagger} \otimes | \beta_{i}\rangle \langle \beta_{j}|.  \nonumber 
\end{align} 
\begin{widetext} 
\begin{align}
\mathcal {F}(\rho ,\Pi ^{a}(\rho))=tr(\rho \Pi ^{a}(\rho))&=
\sum_{ii^{'},jj^{'}}\sqrt{\lambda_{i}\lambda_{i}^{'}\lambda_{j}\lambda_{j}^{'}}\sum_{k}\langle \alpha_{k}| U^{\dagger}|\alpha_{j}\rangle \langle\alpha_{j^{'}}| U|\alpha_{k}\rangle \langle \alpha_{i^{'}}| U |\alpha_{k}\rangle \langle \alpha_{k}| U^{\dagger}|\alpha_{i}\rangle \otimes  \langle \beta_{i^{'}} | \beta_{j}\rangle \langle \beta_{j^{'}} |\beta_{i}\rangle. \label{eq:fidelity}
\end{align}
\end{widetext} 
Since $\rho$ is pure, the fidelity between pre- and post- measurement state is given in eq.(\ref{eq:fidelity}).

After a straight forward calculation, the fidelity between the pre- and post-measurement state is given by
\begin{align}
 \mathcal{F}(\rho,\Pi^{a}(\rho))= \sum_{k} \left(\langle \alpha_{k}| U^{\dagger}\rho^{a} U| \alpha_{k}\rangle \right)^{2} \nonumber
\end{align}
Then, the fidelity based MIN can be written as,
\begin{align}
N_{\mathcal{F}}\left(| \Psi \rangle \langle \Psi|\right)= 1-^{max}_{\Pi^{a}} \sum_{k} \left(\langle \alpha_{k}| U^{\dagger}\rho^{a} U| \alpha_{k}\rangle \right)^{2} \nonumber
\end{align}
where the optimization is over all possible projective measurements. Since the term in the summation is the square of eigenvalues of $\rho^{a}$, we have
\begin{align}
N_{\mathcal{F}}\left(| \Psi \rangle \langle \Psi|\right)= 1-\sum_{k}\lambda^{2}_{k} \nonumber
\end{align}
which is identical with the MIN based on Hilbert-Schmidt norm and skew information.
\section{MIN for arbitrary mixed state}
In order to probe the nonlocal effects due to projective measurement in general bipartite state, we first recall some basic notations and definitions. Let $\{X_{i}:i=0,1,2,\cdots,m^{2}-1\} $ and $\{Y_{j}:j=0,1,2,\cdots,n^{2}-1\} $ be a set of two orthonormal bases corresponding to the Hilbert spaces $\mathcal{H}^{a}$ and $\mathcal{H}^{b}$ respectively, in the sense of $tr(X_{k}X_{l})=tr(Y_{k}Y_{l})=\delta _{kl}$, with $X_{0}=\mathds{1}/\sqrt{m}$ and $Y_{0}=\mathds{1}/\sqrt{n}$. An arbitrary bipartite state in the composite state space $\mathcal{H}^{a}\otimes \mathcal{H}^{b}$ is then defined as
\begin{align}
\rho= \sum_{ij}\gamma _{ij}X_{i}\otimes  Y_{j} \label{SS}
\end{align} 
with $\gamma _{ij} =tr (\rho X_{i}\otimes  Y_{j})$ and $\Gamma =(\gamma _{ij})$ is a correlation matrix with real entries. For any orthonormal basis $\{| k\rangle :k=0,1,2,\cdots,m-1\} $, $| k\rangle \langle k| =\sum_{i}a_{ki}X_{i}$ with $a_{ki}=tr(| k\rangle \langle k|X_{i})$.

\begin{figure*}[!ht]
\centering\includegraphics[width=0.48\linewidth]{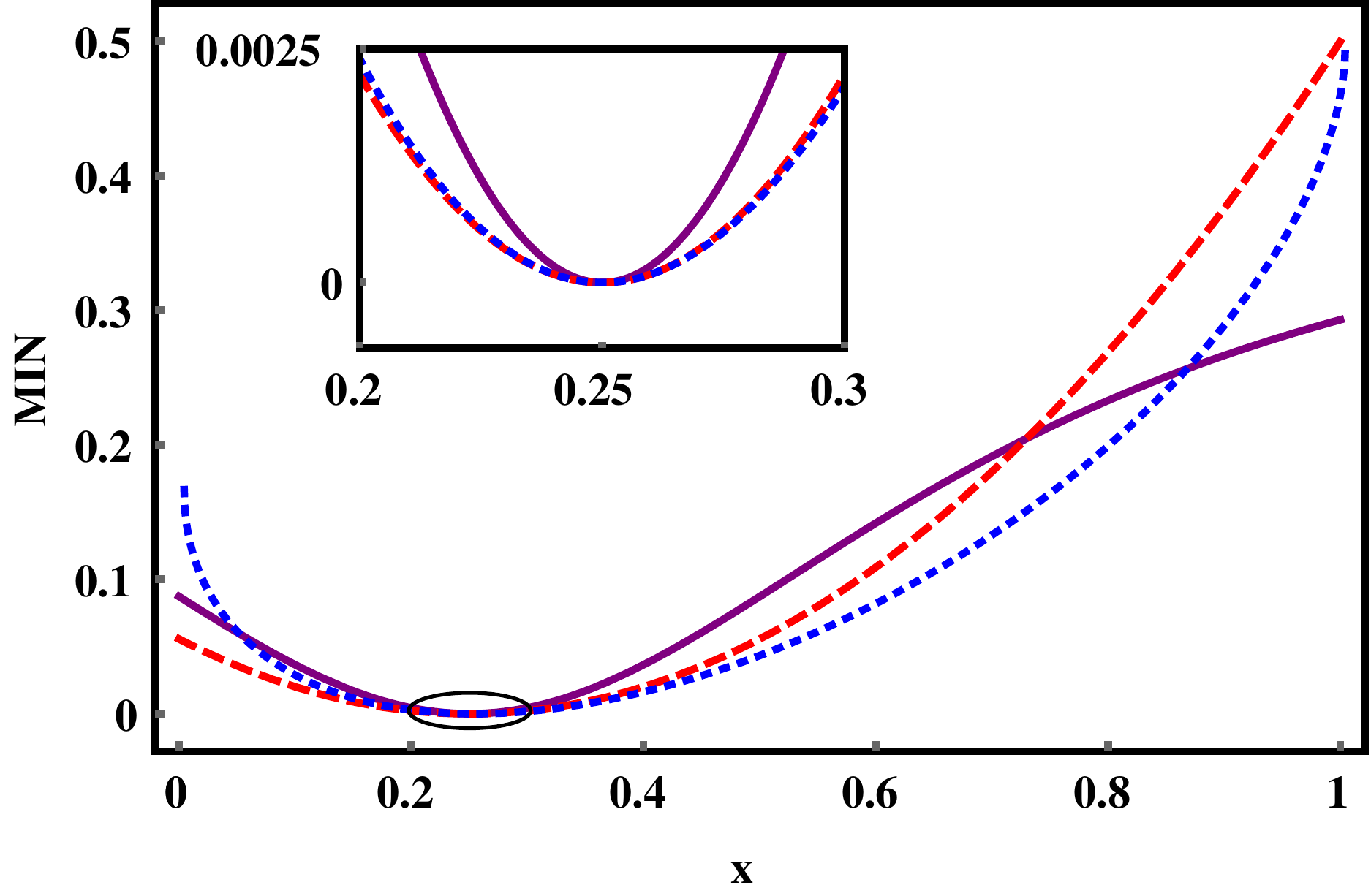}
\centering\includegraphics[width=0.48\linewidth]{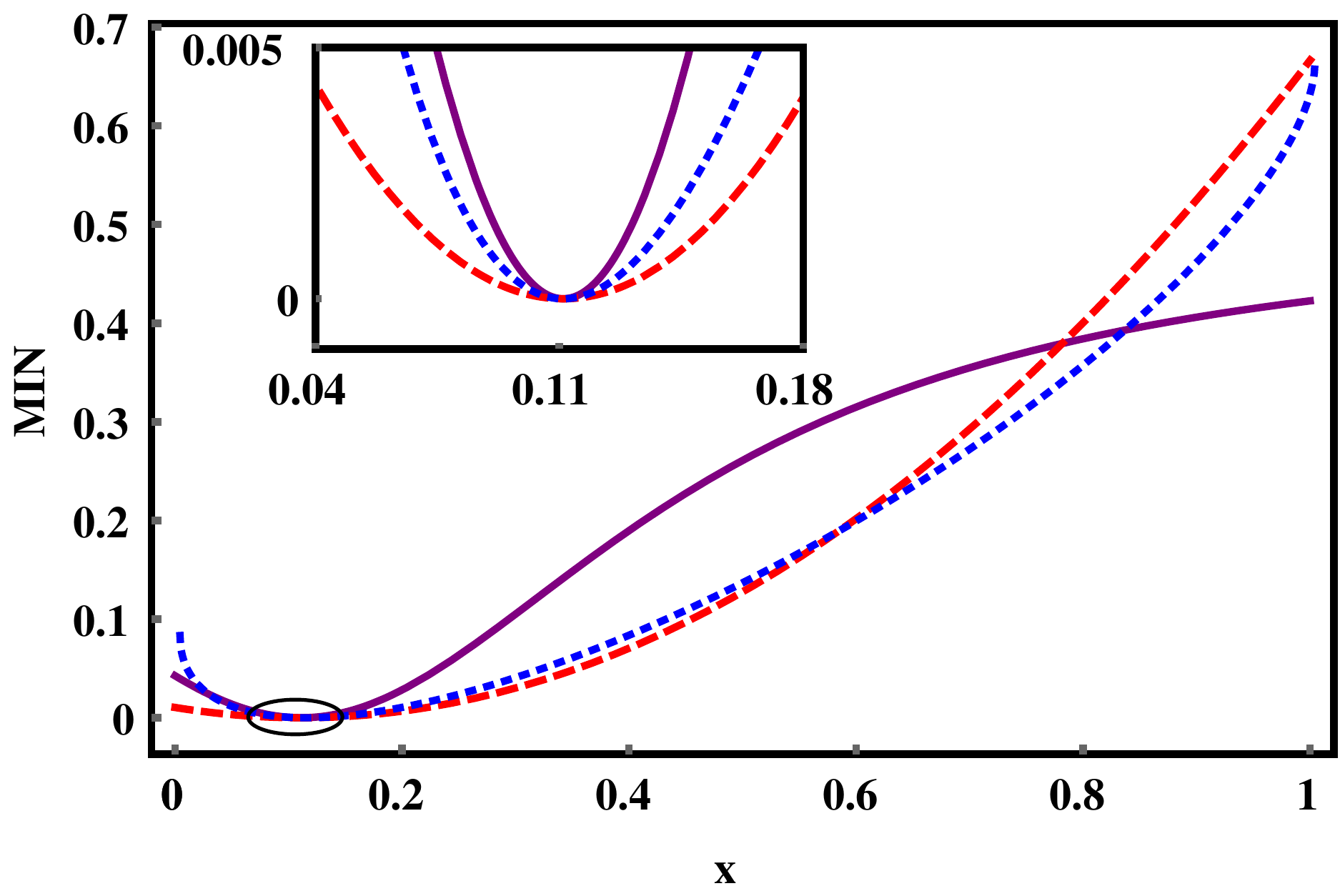}
\caption{(color online) The Hilbert-Schmidt norm (dashed), skew information (dotted) and fidelity (solid) based measurement induced nonlocality of isotropic state for $m=2$ (left) and $m=3$ (right). Insert shows the point at which all forms of MIN vanish identically.}
\label{fig1}
\end{figure*}
\begin{figure*}[!ht]
\centering\includegraphics[width=0.48\linewidth]{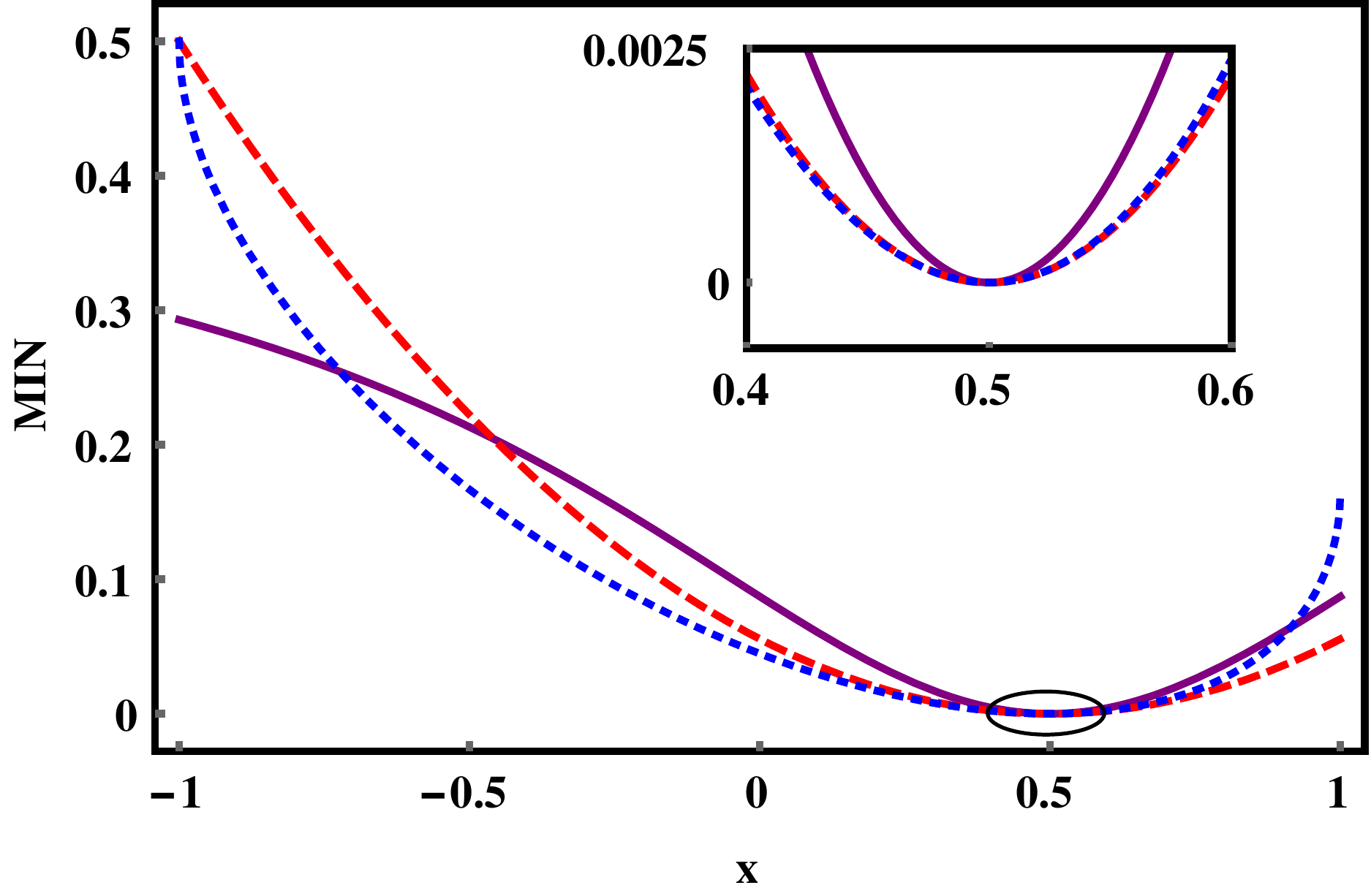}
\centering\includegraphics[width=0.48\linewidth]{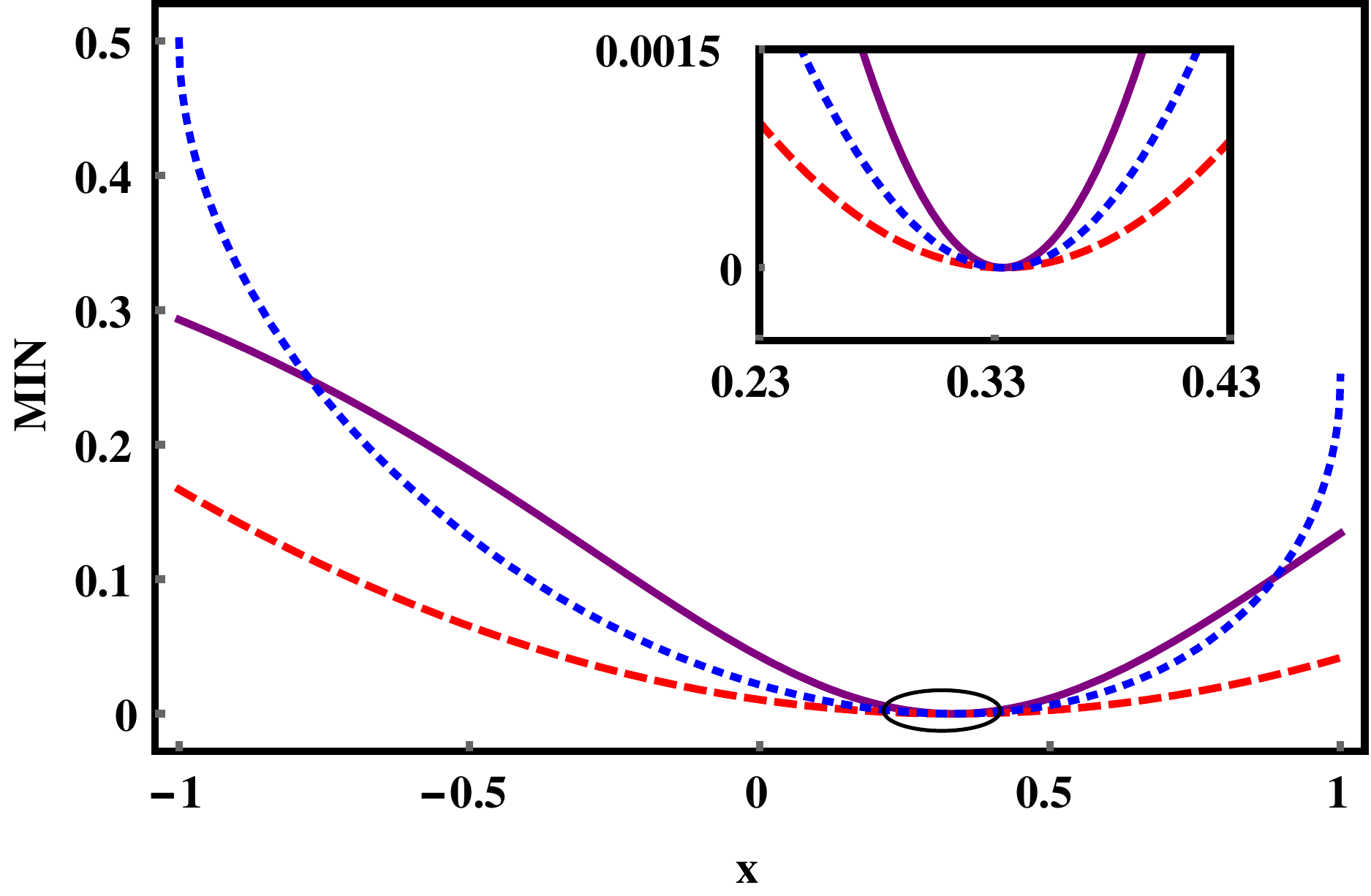}
\caption{(color online) The Hilbert-Schmidt norm (dashed), skew information (dotted) and fidelity (solid) based measurement induced nonlocality of Werner state for $m=2$ (left) and $m=3$ (right). Insert shows the point at which all forms of MIN vanish identically.}
\label{fig2}
\end{figure*}

{\bf Theorem 2:} For any arbitrary bipartite state eq. (\ref{SS}), MIN based on fidelity metric has a tight upper bound as
\begin{align}
N_{\mathcal{F}}(\rho)\leq \frac{1}{\|\Gamma  \|^{2} }\left(\sum_{i=m}^{m^{2}-1}\mu _{i}\right) 
\end{align}
where $\mu _{i}$ are eigenvalues of the matrix $\Gamma \Gamma ^{t}$ listed in increasing order and the superscript denotes transpose of a matrix.

After a straight forward calculation, fidelity between pre- and post-measurement state is computed as
\begin{align}
\mathcal{F}(\rho,\Pi^{a}(\rho) )=\frac{tr(A\Gamma \Gamma ^{t}A^{t})}{\| \Gamma \|^{2} } \nonumber
\end{align} 
where the matrix $A=(a_{ki})$ is a rectangular matrix of order $m\times m^{2}$. Then, MIN is 
\begin{align}
N_{\mathcal{F}}(\rho)=\frac{1}{\|\Gamma  \|^{2} }\left[\|\Gamma  \|^{2}-^{min}_{A}tr(A\Gamma \Gamma ^{t} A^{t})\right]. \label{result}
\end{align}
Now we have,
\begin{align}
\sum_{i=0}^{m^{2}-1}a_{ki}a_{k^{'}i}=tr\left(| k\rangle \langle k| k^{'}\rangle \langle k^{'}| \right)=\delta _{kk^{'}} \nonumber
\end{align}
with $a_{k0}=1/\sqrt{m}$. For $k=k^{'}$
\begin{align}
\sum_{i=1}^{m^{2}-1}a_{ki}^{2}= \frac{m-1}{m} \label{equalk} 
\end{align}
and for $k\neq k^{'}$
\begin{align}
\sum_{i=1}^{m^{2}-1}a_{ki}a_{k^{'}i}= -\frac{1}{m}.  \label{notequalk}
\end{align}
From eq. (\ref{equalk}) and (\ref{notequalk}) we can write the matrix $AA^{t}$~as
\begin{align}\nonumber
AA^{t}=\frac{1}{m}
\begin{pmatrix}
m-1 &  -1 & \cdots & -1 \\
-1 &  m-1 & \cdots & -1 \\
\vdots & \vdots & \ddots & \vdots \\
-1 &  -1 & \cdots & m-1 
\end{pmatrix}
\end{align}
which is a square matrix of order $m$ with eigenvalues $0$ and $1$ (with multiplicity of $m-1$). For this symmetric matrix, we have the similarity transformation $AA^{t}=U D U^{t}$ with real unitary operator $U$ and diagonal matrix $D$. Now constructing $m \times m^{2}$ matrix $B$ as
\begin{align}\nonumber
B=U^{t}A=
\begin{pmatrix}
R \\
0 
\end{pmatrix}
\end{align}
where $R$ is a $(m-1) \times m^{2}$ matrix, such that $RR^{t}=\mathds{1}_{m-1}$ we have
\begin{align}
^{min}_{A}~tr~(A\Gamma \Gamma ^{t}A^{t})= ^{min}_{R}~tr~(R\Gamma \Gamma ^{t}R^{t}).
\end{align}
Then
\begin{align}
N_{\mathcal{F}}(\rho)=\frac{1}{\|\Gamma  \|^{2} }\left[\|\Gamma  \|^{2}-^{min}_{R}tr(R\Gamma \Gamma ^{t} R^{t})\right]. \nonumber
\end{align}
Since
\begin{align}
^{min}_{R:RR^{t}=\mathds{1}_{m-1} }tr(R\Gamma \Gamma ^{t} R^{t}) = \sum_{i=1}^{m-1}\mu _{i}, \nonumber
\end{align}
where $\mu _{i}$ are eigenvalues of the matrix $\Gamma \Gamma ^{t}$ listed in increasing order, we have the following tight bound
\begin{align}
N_{\mathcal{F}}(\rho)\leq \frac{1}{\|\Gamma  \|^{2} }\left(\sum_{i=m}^{m^2-1}\mu _{i}\right)\nonumber
\end{align}
which completes the proof.

{\bf Theorem 3:} For $2\times n$ dimensional systems MIN has the following closed formula,
\begin{eqnarray}
N_{\mathcal{F}}(\rho)=  
\begin{cases}
\frac{1}{\| \Gamma  \|^{2}}(\| \Gamma  \|^{2}-\mu _{1}) & \text{if}~~~ \textbf{x}=0 \\ \frac{1}{\| \Gamma  \|^{2}}(\| \Gamma  \|^{2}-\epsilon ) & \text{if}~~~\textbf{x}\neq0
\end{cases}
\end{eqnarray}
here $\epsilon =tr(A\Gamma \Gamma ^{t}A^{t})$ and 
\begin{eqnarray} \label{eq:ope}
A=\frac{1}{\sqrt{2}}
\begin{pmatrix}
1 & \frac{\textbf{x}}{\| \textbf{x} \|}\\
1 & -\frac{\textbf{x}}{\| \textbf{x} \|}
\end{pmatrix}
\end{eqnarray} 
with $a_{ki}=\langle k| X_{i}|k \rangle $ $(i=0,1,2,3)$.

Proof: Noting that the marginal state $\rho ^{a}=\frac{\mathds{1}^{a}}{2}+\sum_{i=1}^{3}x_{i}X_{i}$ is nondegenerate if and only if $\textbf{x}\neq 0$, the von Neumann projective measurements leaves the marginal states invariant. The eigenprojections are
\begin{eqnarray}
\Pi ^{a}_{1}=\frac{\mathds{1}^{a}}{2}+\frac{\sum_{i=1}^{3}x_{i}X_{i}}{\sqrt{2}\| \textbf{x} \|},~~~~~ \Pi ^{a}_{2}=\frac{\mathds{1}^{a}}{2}-\frac{\sum_{i=1}^{3}x_{i}X_{i}}{\sqrt{2}\| \textbf{x} \|}.\nonumber
\end{eqnarray}
Hence,
 \begin{eqnarray}
a_{1i}=tr(\Pi ^{a}_{1}X_{i})=\frac{x_{i}}{\sqrt{2}\| \textbf{x} \|}=-a_{2i}\label{eq:ele}
\end{eqnarray}
Then using eq.(\ref{eq:ope}) and eq.(\ref{eq:ele}), evaluating eq.(\ref{result}), we will obtain the second equation of Theorem 3. For $\textbf{x}=0$, the state $\rho ^{a}=\frac{\mathds{1}^{a}}{2}$ is degenerate and we compute
\begin{eqnarray}
^{min}_{R:RR^{t}=\mathds{1}_{m-1} }tr(R\Gamma \Gamma ^{t} R^{t}) = \mu _{1}, \nonumber
\end{eqnarray}
which completes the proof of theorem 3.
 
\section{Examples} 
Here we evaluate MIN for two well-known families of mixed state such as isotropic and Werner states. Further, $N_{\mathcal{F}}(\rho)$ is compared with other form of MINs based on skew information \cite{Li2016} and Hilbert-Schmidt norm \cite{Luo2010}.

1. First we consider $m\times m$ dimensional isotropic state in the form \cite{Horodecki1999}
\begin{align}
\rho ^{ab}=\frac{1-x}{m^{2}-1}\mathds{1}+\frac{m^{2}x -1}{m^{2}-1}| \Psi^{+} \rangle \langle \Psi^{+} | \nonumber
\end{align} 
where $| \Psi^{+}\rangle=\frac{1}{\sqrt{m}}\sum_{i}|ii \rangle$, $\mathds{1}$ is identity matrix of order $m^{2}\times m^{2}$ and $x\in [0,1]$. The fidelity based MIN for this state is computed as
\begin{align}
N_{\mathcal{F}}(\rho ^{ab})=\frac{\frac{1}{m}(m^{2} x-1)^{2}}{m(1-x)^{2}+\frac{m-1}{m}(1+m x)^{2}+\frac{1}{m}(m^{2}x-1)^{2}}\nonumber
\end{align}

This result shows that the MIN vanishes only when  $x=1/m^{2}$, at which $\rho^{ab}=\mathds{1}/m^{2}$ being the maximally mixed state.

This result is plotted in Fig.~\ref{fig1} and compared with other forms of MIN for $m=2,3$. It is clearly seen that fidelity based MIN and other forms of MIN show qualitatively similar behaviour. We also note that all the forms of MIN vanish only at $x=\frac{1}{4}\left(\frac{1}{9}\right)$ for $m=2(3)$. Thus the various forms of MIN presented here capture the nonlocal attributes of a quantum state induced by locally invariant measurement consistently.

2. Next we consider $m\times m$ dimensional Werner state \cite{Werner1995}
\begin{align}
\omega ^{ab}=\frac{m-x}{m^{3}-m}\mathds{1}+\frac{m x -1}{m^{3}-m} F \nonumber
\end{align}
where $F=\sum_{\alpha ,\beta }| \alpha \rangle \langle \beta | \otimes  |\beta\rangle \langle \alpha|$ is flip operator with $x\in [-1,1]$. The MIN based on fidelity is computed as,
\begin{align}
N_{\mathcal{F}}(\omega  ^{ab})=\frac{(m x-1)^{2}}{(m-x)^{2}+(m-1)(x-1)^{2}+(m x-)^{2}}\nonumber
\end{align}
This result shows that the MIN vanishes only when $x=1/m$, at which $\omega  ^{ab}=\mathds{1}/m^{2}$ being the maximally mixed state.

This quantity is plotted in Fig.~\ref{fig2} with the other two forms of MIN for $m=2,3$. Here also it is observed that the fidelity based MIN shows similar behaviour as that of MIN in terms skew information and Hilbert-Schmidt norm. We also note that the three different forms of MIN vanish at $x=\frac{1}{2}\left(\frac{1}{3}\right)$ for $m=2(3)$. Here also the consistency of various forms of MIN in
capturing the nonlocal attributes of a quantum state induced by locally invariant measurement is evident.

We also note from Fig.~\ref{fig2} that the range of Hilbert-Schmidt norm based MIN decreases with dimension $m$, unlike the other companion quantities. In fact, we observe that as $m\rightarrow \infty $ the variation in dashed curve with respect to $x$ is tending to zero. On the other hand, the range of other companion quantities are found to be robust with the increase of $m$. Hence the fidelity and skew information based MIN are found to be more sensitive than the Hilbert-Schmidt norm version of MIN in higher dimension.
\section{Conclusions}
In this article, we have proposed a new form of measurement induced nonlocality (MIN) using fidelity induced metric. It is shown that, in addition to capturing global nonlocal effect of a state due to von Neumann projective measurements, this quantity can be remedying local ancilla problem of MIN based on Hilbert-Schmidt norm. We have presented a closed formula of MIN for an arbitrary pure state and $2\times n$ dimensional mixed state. Further we provide an upper bound of fidelity based MIN for $m\times n$ dimensional system. Finally, we have also computed the proposed MIN for the familiar families of states namely, isotropic and Werner states, and showed that they are consistent with the Hilbert-Schmidt norm and skew information based MIN.

\end{document}